\documentclass[conference]{IEEEtran}
\usepackage{cite}
\ifCLASSINFOpdf
   \usepackage[pdftex]{graphicx}
\else
\fi
\usepackage{amsmath,amssymb,amsthm}
\usepackage{steinmetz}
\usepackage{nicefrac}
\usepackage{soul}
\usepackage{multirow}

\usepackage{tikz}
\usepackage{pgfplots}
\usepgfplotslibrary{units}

\usepackage{array}
\ifCLASSOPTIONcompsoc
  \usepackage[caption=false,font=normalsize,labelfont=sf,textfont=sf]{subfig}
\else
  \usepackage[caption=false,font=footnotesize]{subfig}
\fi

\IEEEoverridecommandlockouts

\newcommand{\fixme}[2]{\ifx&#2&{\color{red}#1}\else{\color{red}FIXME\{}#1{\color{red}\}}\footnote{{\color{red}#2}}\PackageWarning{Fixme}{#1: #2}\fi}
\usepackage{booktabs}
\usepackage{color}

\definecolor{Set1-7-1}{RGB}{228,26,28}
\definecolor{Set1-7-2}{RGB}{55,126,184}
\definecolor{Set1-7-3}{RGB}{77,175,74}
\definecolor{Set1-7-4}{RGB}{152,78,163}
\definecolor{Set1-7-5}{RGB}{255,127,0}
\definecolor{Set1-7-6}{RGB}{166,86,40}
\definecolor{Set1-7-7}{RGB}{0,0,0}

\newcommand{\figurewidth}{0.9}	% Width of plots as a fraction of the column width
\newcommand{\figureheight}{0.72}	% Height of plots as a fraction of the column width

\begin{document}

\bstctlcite{IEEEexample:BSTcontrol}

\title{LoRa Symbol Error Rate Under Non-Chip- and Non-Phase-Aligned Interference}
\author{\IEEEauthorblockN{Orion Afisiadis, Matthieu Cotting, Andreas Burg, and Alexios Balatsoukas-Stimming}\\
\IEEEauthorblockA{Telecommunication Circuits Laboratory\\
\'{E}cole polytechnique f\'{e}d\'{e}rale de Lausanne, Switzerland\\
Email: orion.afisiadis@epfl.ch}
}

\maketitle

\begin{abstract}
In this work, we examine the performance of the LoRa chirp spread spectrum modulation in the presence of both additive white Gaussian noise and interference from another LoRa user. To this end, we extend an existing interference model to the more realistic case where the interfering user is neither chip- nor phase-aligned with the signal of interest and we derive an expression for the SER. We show that the existing interference model overestimates the effect of interference on the error rate. Moreover, we derive a low-complexity approximate formula that can significantly reduce the complexity of computing the symbol error rate compared to the complete expression.
\end{abstract}

\IEEEpeerreviewmaketitle

\section{Introduction} \label{sec:intro}
%The Internet of Things (IoT) will consist of billions of connected devices that will enable a large number of novel applications, such as smart city applications, metering, logistics~\cite{Adelantado2017}, and localization and tracking~\cite{Bakkali2017}. Other potential uses of IoT devices include health monitoring~\cite{Catherwood2018}, massive sensor networks for smart farming and environmental monitoring~\cite{Jawad2017}, as well as connected and autonomously driven vehicles that can make road transportation safer and more efficient~\cite{Sharma2018}. Since these devices will mostly be low-power and they are expected to connect wirelessly with each other (or with centralized gateways), several specialized communications protocols have been proposed for IoT applications. Some examples include Sigfox, Weightless, NB-IoT and LoRa~\cite{Raza2017},~\cite{Goursaud2015}.

LoRa is a low-rate, low-power, and high-range modulation that uses chirp spread-spectrum for its physical layer~\cite{SX127x}. LoRa supports multiple spreading factors, coding rates, and packet lengths, to support a wide range of operating signal-to-noise ratios (SNRs). The LoRa physical layer is proprietary, but reverse engineering attempts that have been carried out~\cite{Knight2016,Robyns2018} have revealed the mathematical description~\cite{Vangelista2017}. The effect of carrier- and sampling-frequency offset on LoRa digital receivers has been modeled and analyzed in~\cite{Ghanaatian2019}. For MAC layer, LoRa relies on LoRaWAN, which uses an ALOHA-based mechanism, meaning that collisions are not explicitly avoided and that the scalability of the network is potentially limited by inter-node interference~\cite{Bor2016,Haxhibeqiri2018}. An overview and performance evaluations of LoRaWAN can be found in~\cite{Haxhibeqiri2018}.

Since LoRa uses the ISM band, interference from other technologies using the same band is another potential problem and has received some attention in the literature. Specifically,~\cite{Orfanidis2017} studies the co-existence of LoRa with IEEE 802.15.4g, while~\cite{Reynders2016} studies the co-existence of LoRa with ultra-narrowband technologies, such as Sigfox. The impact of interference coming from other LoRa nodes has also received some attention. Specifically, the work of~\cite{Croce2018} examines the effect of imperfect orthogonality between different LoRa spreading factors by examining the signal-to-interference ratio (SIR) threshold for receiving a packet correctly for all combinations of spreading factors. However, interference is particularly detrimental when users with the same spreading factor collide. The authors of~\cite{Feltrin2018} perform an experimental assessment of the link-level characteristics of the LoRa system, followed by a system-level simulation to assess the capacity of a LoRaWAN network. Convenient approximate formulas for the bit-error rate (BER) of the LoRa modulation when transmission takes place over additive white Gaussian noise (AWGN) and Rayleigh fading channels are given in~\cite{Elshabrawy2018} but collisions are not considered. Finally, the work of~\cite{Elshabrawy2018b}, which is most closely related to our work, provides an approximation for the BER of the LoRa modulation under AWGN and interference from a single LoRa interferer with the same spreading factor (\emph{same-SF}). Capacity planning for LoRa with the aforementioned interference model is addressed in~\cite{Elshabrawy2019}.

\subsubsection*{Contributions}
The work of~\cite{Elshabrawy2018b} made a significant first step toward understanding of the behavior of LoRa under same-SF interference. In this work, we extend the interference model of~\cite{Elshabrawy2018b} to the more general (and more realistic) case where the interference is neither chip- nor phase-aligned with the signal-of-interest. We derive an expression for the symbol-error rate (SER) under this new complete interference model. Moreover, we derive an approximation for the SER under the new interference model and we show that non-integer chip duration time-misalignment in particular has a significant effect on the SER. Specifically, we show that the interference model of~\cite{Elshabrawy2018b} is pessimistic in the sense that it consistently over-estimates the actual SER.

\section{The LoRa Modulation} \label{sec:system_model}
LoRa is a spread-spectrum modulation that uses a bandwidth $B$ and $N = 2^\text{SF}$ chips per symbol, where SF is called the \emph{spreading factor} with $\text{SF} \in \{7, \dots, 12\}$. When considering the discrete-time baseband equivalent signal, the bandwidth $B$ is split into $N$ frequency steps. A symbol $s \in \mathcal{S}$, where $ \mathcal{S} = \left\{0,\hdots,N{-}1\right\} $, begins at frequency $(\frac{s B}{N} - \frac{B}{2})$. The frequency increases by $\frac{B}{N}$ at each chip until it reaches the Nyquist frequency $\frac{B}{2}$. When the Nyquist frequency is reached, there is a frequency fold to $-\frac{B}{2}$ at chip $n_{\text{fold}} = N-s$. The general discrete-time baseband equivalent description of a LoRa symbol $s$ is
\begin{align}
x_s[n] & =
     \begin{cases}
       e^{j2\pi\left( \frac{1}{2N} \left(\frac{B}{f_s}\right)^2n^2 + \left(\frac{s}{N} - \frac{1}{2}\right)\left(\frac{B}{f_s}\right)n \right)}, &  n \in \mathcal{S}_1\\
       e^{j2\pi\left( \frac{1}{2N} \left(\frac{B}{f_s}\right)^2n^2 + \left(\frac{s}{N} - \frac{3}{2}\right)\left(\frac{B}{f_s}\right)n \right)}, &  n \in \mathcal{S}_2,
     \end{cases}
\end{align}
where $\mathcal{S}_1 = \{0,..., n_{\text{fold}} -1\}$ and $\mathcal{S}_2 = \{n_{\text{fold}},...,N-1\}$. In the practically relevant case where the sampling frequency $f_s$ is equal to $B$, which we assume for the remainder of this manuscript, the discrete-time baseband equivalent equation of a LoRa symbol $s$ can be simplified to
\begin{align} \label{eq:LoRa_symbol}
x_s[n] & = e^{j2\pi \left(\frac{n^2}{2N}  + \left(\frac{s}{N} - \frac{1}{2}\right)n \right)}, \;\; n \in \mathcal{S}.
\end{align}
After transmission over a time-invariant and frequency-flat wireless channel with complex-valued channel gain $h \in \mathbb{C}$, the received LoRa symbol is given by
\begin{align}
  y[n] & = hx_s[n] + z[n], \;\; n \in \mathcal{S}, \label{eq:lora_rx}
\end{align}
where $z[n] \sim \mathcal{CN}(0,\sigma^2)$ is complex additive white Gaussian noise with $\sigma^{2} = \frac{N_0}{N}$ and $N_0$ is the singled-sided noise power spectral density. We assume that $|h| = 1$ without loss of generality, so that the signal-to-noise ratio (SNR) is $\frac{1}{N_0}$.

To demodulate the symbols, first a \emph{dechirping} is performed, where the received signal is multiplied by the complex conjugate of a reference signal $x_{\text{ref}}$. A convenient choice for this reference signal is an \emph{upchirp}, i.e., the LoRa symbol for $s=0$
\begin{align}
  x_{\text{ref}}[n] = e^{j2\pi \left(\frac{n^2}{2N} - \frac{n}{2} \right)}, \;\; n \in \mathcal{S}. \label{eq:Ref_symbol}
\end{align}
Then, the non-normalized discrete Fourier transform (DFT) is applied to the dechirped signal in order to obtain $\mathbf{Y} = \text{DFT}\left(\mathbf{y} \odot \mathbf{x}_{\text{ref}}^{*}\right)$, where $\odot$ denotes the Hadamard product and $\mathbf{y} = \begin{bmatrix} y[0] & \hdots & y[N-1] \end{bmatrix}$ and $\mathbf{x}_{\text{ref}} = \begin{bmatrix} x_{\text{ref}}[0] & \hdots & x_{\text{ref}}[N-1] \end{bmatrix}$. Demodulation can be performed by selecting the bin index with the maximum magnitude
\begin{align}
  \hat{s} = \arg\max_{k\in \mathcal{S} } \left( |Y_k| \right). \label{eq:retrieved_symbol_dft}
\end{align}

\section{Symbol Error Rate Under AWGN} \label{sec:SER_AWGN}

In this section, we first derive the expression for the LoRa SER under additive white Gaussian noise (AWGN), which is useful for later explaining how the SER can be calculated in the presence of both AWGN and interference.

\subsection{Distribution of the Decision Metric}
In the absence of noise, and with perfect synchronization, the DFT of the dechirped signal $\mathbf{Y}$ has a single frequency bin that contains all the signal energy (i.e., a bin with magnitude $N$) and all remaining $N-1$ bins have zero energy. On the other hand, when AWGN is present, all frequency bins will contain some energy. The distribution of the frequency bin values $Y_k$ for $k \in \mathcal{S}$ is
\begin{align}
  Y_k \sim
  \begin{cases}
    \mathcal{CN}\left(0,\sigma ^2\right), & k \in \mathcal{S} / s \\
    \mathcal{CN}\left(N(\cos \phi + j \sin \phi),\sigma ^2\right), & k = s, \\
  \end{cases}
\end{align}
where $\phi = \phase{h}$ denotes a phase shift introduced by the transmission channel $h$ that is fixed for each transmission but generally uniformly distributed in $[0,2\pi)$ and $s$ is the transmitted symbol.

Let us define $Y'_k = \frac{Y_k}{\sigma}$ for $k \in \mathcal{S}$. The values $Y'_k$ can be used in~\eqref{eq:retrieved_symbol_dft} instead of $Y_k$ without changing the result and their distribution is
\begin{align}\label{eq:distribution_Yk'}
  Y'_k \sim
  \begin{cases}
    \mathcal{CN}\left(0,1 \right), & k \in \mathcal{S} / s \\
    \mathcal{CN}\left(\frac{N\cos \phi}{\sigma} + j\frac{N\sin \phi}{\sigma},1 \right), & k = s. \\
  \end{cases}
\end{align}
Using basic properties of the complex normal distribution, we can show that the demodulation metric $|Y'_k|$ follows a Rayleigh distribution for $k \in \mathcal{S} / s$ and a Rice distribution for $k=s$
\begin{align}
  |Y'_k| \sim
  \begin{cases}
    f_{\text{Ra}}(y;1), & k \in \mathcal{S} / s \\
    f_{\text{Ri}}\left(y;\frac{N}{\sigma},1\right), & k = s.
  \end{cases}
\end{align}
We denote the probability density function (PDF) and the cumulative density function (CDF) of the Rayleigh and Rice distributions by $f_{\text{Ra}}(y; \sigma)$, $f_{\text{Ri}}(y; v, \sigma)$ and $F_{\text{Ra}}(y;\sigma)$, $F_{\text{Ri}}(y;v, \sigma)$, respectively, where $\sigma$ and $v$ denote the \emph{scale} and \emph{location} parameters.

\subsection{Symbol Error Rate}
A symbol error occurs if and only if any of the $|Y'_k|$ values for $k \in \mathcal{S}/s$ exceeds the value of $|Y'_s|$, or, equivalently, if and only if $|Y'_{\max}| > |Y'_s|$, where $|Y'_{\max}| = \max _{k \in \mathcal{S}/s}|Y'_k|$. Using order statistics~\cite{David2003} and the fact that all $|Y'_k|$ for $k \in \mathcal{S}/s$  are i.i.d., the PDF of $|Y'_{\max}|$ can be obtained as
\begin{align} \label{eq:2_to_sf_th_order_pdf}
  f_{|Y'_{\max}|} (y) = \left(N-1\right) f_{\text{Ra}}(y;1) F_{\text{Ra}}\left(y;1\right)^{(N-2)}
\end{align}
Using $f_{|Y'_{\max}|} (y)$, the conditional SER when symbol $s$ is transmitted can be calculated as
\begin{align} \label{eq:formula_SER_only_noise}
  \small
  P(\hat{s} \neq s|s)  & = \int_{y=0}^{+ \infty} \int_{x=0}^{y} f_{\text{Ri}}\left(x;v,1\right) f_{|Y'_{\max}|} (y) {dx} {dy} \\
              & = \int_{y=0}^{+\infty} F_{\text{Ri}}\left(y;v,1\right) f_{|Y'_{\max}|} (y) dy, \label{eq:formula_SER_only_noise_finalform}
\end{align}
with $v = \frac{N}{\sigma}$. The SER for all symbols $s$ is identical, meaning that~\eqref{eq:formula_SER_only_noise_finalform} is in fact equal to the average SER and, if we assume that all symbols are equiprobable, also the expected SER.

\subsection{Symbol Error Rate Approximations}
While the evaluation of~\eqref{eq:formula_SER_only_noise_finalform} is in principle straightforward, in practice the values of $N$ in the LoRa modulation are very large so that numerical problems arise. For this reason, two approximations that can be used to efficiently evaluate~\eqref{eq:formula_SER_only_noise_finalform} were derived in~\cite{Elshabrawy2018}. Specifically,~\cite{Elshabrawy2018} used a Gaussian approximation so that $|Y'_s| \sim \mathcal{N}\left(\frac{N}{\sigma},1\right)$ and $|Y'_{\max}| \sim \mathcal{N}\left(\mu_{\beta},\sigma^2_{\beta}\right)$ and where appropriate expressions are given to calculate $\mu_{\beta}$ and $\sigma^2_{\beta}$. With our definition of the SNR, the SER can be calculated as
\begin{align}
  P(\hat{s} \neq s) & \approx Q\left(\frac{\sqrt{\text{SNR}} - \left((H_{N-1})^2 - \frac{\pi^2}{12}\right)^{1/4}}{\sqrt{H_{N-1} - \sqrt{(H_{N-1})^2 - \frac{\pi^2}{12}} + 0.5}}\right), \label{eq:SER_approximation}
\end{align}
where $H_n = \sum_{k=1}^n\frac{1}{k}$ denotes the $n$th harmonic number and $Q(\cdot)$ denotes the Q-function.

\begin{figure}
	\centering
	\includegraphics[width=0.45\textwidth]{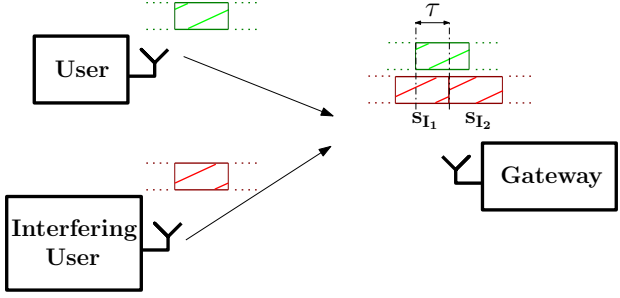}
	\caption{Illustration of LoRa uplink transmission with one interfering user having an arbitrary $\tau$.}
	\label{fig:interf_illustr}
\end{figure}

\section{Symbol Error Rate Under AWGN and \\Same-SF LoRa Interference} \label{sec:SER_interf}

In this section, we analyze the case of a gateway trying to decode the message of a user in the presence of an interfering LoRa device, as depicted in~Fig.~\ref{fig:interf_illustr}. We assume that the LoRa gateway is perfectly synchronized to the user whose message is decoded. Various synchronization techniques for LoRa have been explained in the literature~\cite{Robyns2018}. It has been shown that interferers with different spreading factors can be considered approximately orthogonal~\cite{Adelantado2017,Goursaud2015}. Therefore, in this work we limit our model to interference signals with the same spreading factor as the one employed by the user. Finally, for simplicity, in this work we only consider one interfering user. In this case, the general signal model is
\begin{align}
  y[n] & = hx[n] + h_{I}x_{I}[n] + z[n], \;\; n \in \mathcal{S}, \label{eq:lora_rx_int}
\end{align}
where $h$ is the channel gain between the user of interest and the LoRa gateway, $x[n]$ is the signal of interest, $h_{I}$ is the channel gain between the interferer and the LoRa gateway, $x_{I}[n]$ is the interfering signal, and $z[n] \sim \mathcal{N}(0,\sigma^2)$ is additive white Gaussian noise. Since we assume that $|h| = 1$, the signal-to-interference ratio (SIR) can be defined as
\begin{align}
  \text{SIR}  & = \frac{1}{|h_I|^2} = \frac{1}{P_I},
\end{align}
where we use $P_I$ to denote the power of the interfering user. Since LoRa uses the (non-slotted) ALOHA protocol for medium access control, the interfering signal ${y_{I}[n] = h_{I}x_{I}[n]}$ is not synchronized in any way to the user or the gateway. Due to the lack of synchronization, the interfering signal $x_{I}[n]$ will generally be a combination of parts of two distinct LoRa symbols, which we denote by $s_{I_{1}}$ and $s_{I_{2}}$, as shown in~Fig.~\ref{fig:interf_illustr}.

Let $\tau$ denote the relative time-offset between the first chip of the symbol of interest $s$ and the first chip of the interfering symbol $s_{I_{2}}$ (i.e., the first chip of the interfering symbol $s_{I_{2}}$ starts $\tau$ chip durations \emph{after} the first chip of $s$). Due to the complete lack of synchronization, we assume that $\tau$ is uniformly distributed in $[0, N)$. We note that in~\cite{Elshabrawy2018b}, the offset $\tau$ is constrained to integer chip durations, which is not particularly realistic since it effectively assumes that the interferer is chip-aligned with the user. Let $ \mathcal{N}_{L_1} = \{0, \dots,\lceil \tau\rceil-1\}$ and $ \mathcal{N}_{L_2} = \{\lceil \tau\rceil, \dots, N-1 \} $. The discrete-time baseband equivalent equation of $x_I[n]$ can be found using~\eqref{eq:LoRa_symbol} for $ s_{I_{1}} $ and $ s_{I_{2}}$, appropriately adjusted to include the offset $\tau$
\begin{align}
x_{I}[n] & =
     \begin{cases}
       e^{j2\pi \left(\frac{(n + N-\tau)^{2}}{2N} + (n + N-\tau)\left(\frac{s_{I_{1}}}{N}-\frac{1}{2}\right)\right)}, & n \in \mathcal{N}_{L_1}\\
       e^{j2\pi \left(\frac{(n - \tau)^{2}}{2N} + (n - \tau)\left(\frac{s_{I_{2}}}{N}-\frac{1}{2}\right)\right)}, & n \in \mathcal{N}_{L_2}.
     \end{cases}
\end{align}
The demodulation of $y[n]$ at the receiver yields
\begin{align}
  \mathbf{Y}  & = \text{DFT}\left(\mathbf{y} \odot \mathbf{x}_{\text{ref}}^{*}\right) \\
              & = \text{DFT}\left(h\mathbf{x} \odot \mathbf{x}_{\text{ref}}^{*}\right) + \text{DFT}\left(h_I\mathbf{x_I} \odot \mathbf{x}_{\text{ref}}^{*}\right) + \text{DFT}\left(\mathbf{z} \odot \mathbf{x}_{\text{ref}}^{*}\right).
\end{align}
We call $\text{DFT}\left(\mathbf{x_I} \odot \mathbf{x}_{\text{ref}}^{*}\right)$ and $\text{DFT}\left(h_I\mathbf{x_I} \odot \mathbf{x}_{\text{ref}}^{*}\right) = \text{DFT}(\mathbf{y}_{I} \odot \mathbf{x}_{\text{ref}}^{*})$ the transmitted and received \emph{interference patterns}, respectively. It is clear that the interference pattern depends on the time-domain interference signal $\mathbf{y}_{I}$, which is in turn a function of the interfering symbols $ s_{I_{1}} $, $ s_{I_{2}} $, the channel $h_I$, and the interferer time-offset $ \tau $.

\begin{figure*}
  \begin{align}
    P\left(\hat{s}\neq s\right) & = 1-\frac{1}{2 \pi N^4}\sum_{s=0}^{N-1}\sum _{s_{I_{1}}=0}^{N-1}\sum _{s_{I_{2}}=0}^{N-1} \int _{\tau = 0}^{N}\int_{\omega=0}^{2 \pi} \int_{y=0}^{+\infty} f_{\text{Ri}}\left(y;v_{s},1\right) \prod_{\substack{k=1\\k\neq s}}^{N} F_{\text{Ri}}(y;v_k,1) dyd\omega d \tau. \label{eq:ser_full}
  \end{align}
  \hrule
\end{figure*}

\subsection{Distribution of the Decision Metric}\label{sec:distdecmetr}
Let $ R_{k} $ denote the value of the transmitted interference pattern at frequency bin $k$, i.e.,
\begin{align}
R_{k} =  \text{DFT}\left(\mathbf{x_I} \odot \mathbf{x}_{\text{ref}}^{*}\right)[k], \; k \in \mathcal{S}.
\end{align}
For a specific combination of a symbol s and an interference pattern $\mathbf{y}_I$, adding the interference to the signal of interest corresponds to changing the mean value of the distribution of $ Y'_k $ in~\eqref{eq:distribution_Yk'}, as follows:
{
\small
\begin{align}
  Y'_k {\sim}
  \begin{cases}
    \mathcal{CN}\left(\frac{|h_IR_{k}|\cos\theta}{\sigma} {+} j\frac{|h_IR_{k}|\sin\theta}{\sigma},1 \right), & k \in \mathcal{S} / s \\
    \mathcal{CN}\left(\frac{N\cos\phi {+} |h_IR_{k}|\cos\theta}{\sigma} {+} j\frac{N\sin\phi {+} h_I|R_{k}|\sin\theta}{\sigma},1 \right), & k = s, \\
  \end{cases}
\end{align}
}%
where $\theta = \phase{h_{I}}$ is the phase shift introduced by the interference channel and is fixed for each transmission but generally uniformly distributed in $[0,2\pi)$. Thus, in the presence of interference, the demodulation metric $|Y'_k|$ used in~\eqref{eq:retrieved_symbol_dft} is distributed according to
\begin{align}
  |Y'_k| {\sim}
  \begin{cases}
    f_{\text{Ri}}\left(y;\frac{|h_{I}R_k|}{\sigma},1\right), & k \in \mathcal{S} / s \\
		f_{\text{Ri}}\left(y;\frac{\sqrt{N^2 {+} |h_{I}R_{k}|^2 {+} 2 N|h_{I}R_{k}|\cos(\omega)}}{\sigma},1\right), & k = s, \label{eq:formula_Riceks_noise_interf} \\
  \end{cases}
\end{align}
where we define the phase shift between the user and the interfering user as $\omega = \phi-\theta$ for simplicity.

\subsection{Symbol Error Rate}
Similarly to~\eqref{eq:formula_SER_only_noise_finalform}, in the presence of interference the SER conditioned on $s$, $\mathbf{y}_{I}$, and $\omega$, can be written as
\begin{align} \label{eq:formula_SER_noise_interf}
  \small
	P\left(\hat{s}\neq s|s,\mathbf{y}_{I},\omega\right) & = 1-\int_{y=0}^{+\infty} f_{\text{Ri}}\left(y;v_{s},1\right) F_{|Y'_{\max}|} (y) dy,
\end{align}
where $v_{s} = \frac{1}{\sigma}\sqrt{N^2+ |h_{I}R_{s}|^2 + 2 N|h_{I}R_{s}|\cos(\omega)}$ is the location parameter for the bin $k=s$. The CDF of the $N$th order statistic is known to be $F_{n}(x) = P(X_1 < x) P(X_2 < x) \dots P(X_n < x)$. Thus, we can directly deduce that the CDF of the maximum interfering bin is
\begin{equation}
	F_{|Y'_{\max}|} (y) = \prod_{\substack{k=1\\k\neq s}}^{N} F_{\text{Ri}}(y;v_k,1), \label{eq:cdf_back_bins}
\end{equation}
where $v_k = \frac{|h_{I}R_k|}{\sigma}$.
%
%\begin{figure}
%	\centering
%	\includegraphics[width=0.28\textwidth]{figs/binOfInt}
%	\caption{Vector representation of the desired signal (green), interference (black) and noise (red) at the bin of interest.}
%	\label{fig:binOfInt}
%\end{figure}
%
By taking the expectation of $P\left(\hat{s}\neq s|s,\mathbf{y}_{I},\omega\right)$ with respect to $\omega$, we get the SER conditioned on $s$, $\mathbf{y}_{I}$
\begin{align}
	P\left(\hat{s}\neq s|s,\mathbf{y}_{I}\right) & = \frac{1}{2 \pi} \int_{\omega=0}^{2 \pi} P\left(\hat{s}\neq s|s,\mathbf{y}_{I},\omega\right)d\omega \label{eq:cond_prob}.
\end{align}
Recall that, by assumption, $s_{I_{1}}$ and $s_{I_{2}}$ are uniformly distributed in $\mathcal{S}$ and $\tau$ is uniformly distributed in $[0,N)$. As such, the conditional SER $P\left(\hat{s}\neq s|s\right)$ can be computed as
\begin{align}
  P\left(\hat{s}\neq s|s\right) & = \frac{1}{N^3}\sum _{s_{I_{1}}=0}^{N-1}\sum _{s_{I_{2}}=0}^{N-1} \int _{0}^{N} P\left(\hat{s}\neq s|s,\mathbf{y}_{I}\right)d \tau. \label{eq:cond_avg}
\end{align}
Finally, $s$ is also uniformly distributed in  $\mathcal{S}$, so that the unconditional SER becomes
\begin{align}
  P\left(\hat{s}\neq s\right) & = \frac{1}{N}\sum_{s=0}^{N-1}P\left(\hat{s}\neq s|s\right). \label{eq:final_avg}
\end{align}
The full expression for $P\left(\hat{s}\neq s\right) $ is given in~\eqref{eq:ser_full}.

\section{Symbol Error Rate Approximation} \label{sec:SER_approx}
Apart from the numerical problems that arise from the product of $(N{-}1)$ CDFs in~\eqref{eq:ser_full}, the computational complexity of evaluating~\eqref{eq:ser_full} is very high. For this reason, in this section, we derive an approximation for~\eqref{eq:ser_full}.

\subsection{Interference Patterns}
We first derive an explicit form for the magnitude of the transmitted interference pattern $R_k,~k\in\mathcal{S}$. Note that the offset $\tau$ can be split into an integer part $L$ and a non-integer part $\lambda$, i.e., $L = \left \lfloor{\tau}\right \rfloor,$ and $\lambda = \tau - \left \lfloor{\tau}\right \rfloor$. Using the definition of the DFT and after some algebraic transformations, we obtain
\begin{align}
  R_{k} & = A_{k,1} e^{-j\theta_{k,1}} + A_{k,2} e^{-j\theta_{k,2}}, \label{eq:short_A_theta}
\end{align}%
where
\begin{align}
	A_{k,1} & = \frac{\sin \left( \frac{\pi}{N} (s_{I_{1}}-k-\tau)\lceil \tau \rceil \right)}{\sin \left( \frac{\pi}{N} (s_{I_{1}}-k-\tau) \right)}, \label{eq:A1} \\
  A_{k,2} &=  \frac{\sin \left( \frac{\pi}{N} (s_{I_{2}}-k-\tau)(N-\lceil \tau \rceil) \right)}{\sin \left( \frac{\pi}{N} (s_{I_{2}}-k-\tau) \right)}, \label{eq:A2}
\end{align}
and
\begin{align}
  \theta_{k,1} & = \frac{\pi}{N} \left( {-}\tau^2 + (\lambda{-}L)N + s_{I_{1}}(2\tau -\lceil \tau \rceil +1) + \right. \nonumber\\
             &  + \left. k(\lceil \tau \rceil -1) {+} \tau(\lceil \tau \rceil -1) \right), \\
  \theta_{k,2} & = \frac{\pi}{N} \left( {-}\tau^2 + s_{I_{2}}(2\tau -\lceil \tau \rceil +1-N) + \right. \nonumber\\
   &  + \left. k(\lceil \tau \rceil -1+N) + \tau(\lceil \tau \rceil -1) \right).
\end{align}
We define $[x]_y = x \mod y$. For the special case where $\tau$ is an integer and $k = [s_{I_{1}}-\tau]_N$ and $k = [s_{I_{2}}-\tau]_N$, \eqref{eq:A1} and \eqref{eq:A2}, respectively, are of the indeterminate form $\nicefrac{0}{0}$. Using L'H\^{o}pital's rule, it can be shown that in these cases we have $ A_{k,1} = \lceil \tau \rceil $, and $ A_{k,2} = N - \lceil \tau \rceil $. The magnitude of $R_k$ in~\eqref{eq:short_A_theta} can be written as
\begin{align}
  |R_{k}| & = \sqrt{A_{k,1}^{2} + A_{k,2}^{2} + 2A_{k,1}A_{k,2}\cos(\theta_{k,1} - \theta_{k,2})}. \label{eq:final_Yk}
\end{align}

\subsection{Symbol Error Rate Approximation}
We now follow a procedure that is similar to the procedure in~\cite{Elshabrawy2018b} in order to derive a simple approximation for $P\left(\hat{s}\neq s\right)$ that is also more efficient to evaluate than~\eqref{eq:ser_full}. First, using the triangle inequality, we can simplify~\eqref{eq:final_Yk} to
\begin{align}
  |R_{k}| & \approx  |A_{k,1}|+|A_{k,2}|. \label{eq:triangle}
\end{align}
We assume that the interference-induced SER is dominated by the maximum of $|R_k|$. Thus, we are interested in evaluating
\begin{align}
  |R_{k_{\max}}| & = \max_k\left(|A_{k,1}|+|A_{k,2}|\right).
\end{align}
Similarly to~\cite{Elshabrawy2018b}, we choose
\begin{align}
  k_{\max} & \approx \arg \max_k \left(|A_{k,2}|\right) = \lfloor \tau \rceil,
\end{align}
so that we can easily approximate $R_{\max}$ as
\begin{align}
    |R_{k_{\max}}| & \approx |A_{\lfloor \tau \rceil,1} + A_{\lfloor \tau \rceil,2}|. \label{eq:maxapprox}
\end{align}
The probability that the (maximum) interference bin $\lfloor \tau \rceil$ coincides with the bin of the signal-of-interest $s$ is $\frac{1}{N}$. Thus, for the approximation of the SER, we only consider the cases where $\lfloor \tau \rceil \neq s$. Only considering $\lfloor \tau \rceil \neq s$ also has the convenient side-effect that we ignore the only case of~\eqref{eq:formula_Riceks_noise_interf} which contains $\omega$, meaning that we can entirely avoid the integration over $\omega$ in the computation of $P\left(\hat{s}\neq s|s,\mathbf{y}_{I}\right)$. Let $P^{(I)}(\hat{s}\neq s)$ denote the SER under interference resulting from the approximation in~\eqref{eq:maxapprox}. As explained in Section~\ref{sec:distdecmetr}, $|Y'_{k_{\max}}|$ follows a Rice distribution, which can be approximated by a Gaussian distribution for large location parameters~\cite{Elshabrawy2018b} so that
\begin{align}
  |Y'_{k_{\max}}| \mathrel{\dot\sim} \mathcal{N}\left(\frac{|h_I||R_{k_{\max}}|}{\sigma},1\right).
\end{align}
\begin{figure}[t]
  \centering
  \begin{tikzpicture}

	%\pgfplotsset{grid style={dashed}}
	\small

	\begin{semilogyaxis}[
		width = \figurewidth\columnwidth,
		height = \figureheight\columnwidth,
		xlabel = {SNR (dB)},
		ylabel = {Symbol Error Rate},
		label style={font=\small},
    tick label style={font=\footnotesize},
		ylabel near ticks,
		xlabel near ticks,
		xmin = -25, xmax = 0,
		ymin = 1e-5, ymax = 1,
		grid = both,
		legend image post style={scale=0.7},
		legend style={at={(0.45,-0.2)},anchor=north,font=\tiny},
		legend cell align={left},
		legend columns={7},
		%transpose legend,
	]

		\addlegendimage{empty legend}
    \addlegendentry{\textbf{PO:}} 

		%% Phase offset %%
		% SF = 7
		\addplot[Set1-7-4, thick, solid, mark=star, mark options={scale=1.2}] table[x index=0, y index = 1] {figs/data/RES_SF7_Pi-3.00_OS10_phaseOffset1_symmetries0_0.dat};
		\addlegendentry{SF${=}7$}
		% SF = 8
		\addplot[Set1-7-5, thick, solid, mark=pentagon*, mark options={scale=1.2}] table[x index=0, y index = 1] {figs/data/RES_SF8_Pi-3.00_OS10_phaseOffset1_symmetries0_0.dat};
		\addlegendentry{SF${=}8$}
		% SF = 9
		\addplot[Set1-7-1, thick, solid, mark=*, mark options={scale=1.1}] table[x index=0, y index = 1] {figs/data/RES_SF9_Pi-3.00_OS10_phaseOffset1_symmetries0_0.dat};
		\addlegendentry{SF${=}9$}
		% SF = 10
		\addplot[Set1-7-2, thick, solid, mark=square*, mark options={scale=1.05}] table[x index=0, y index = 1] {figs/data/RES_SF10_Pi-3.00_OS10_phaseOffset1_symmetries0_0.dat};
		\addlegendentry{SF${=}10$}
		% SF = 11
		\addplot[Set1-7-3, thick, solid, mark=triangle*, mark options={scale=1.2}] table[x index=0, y index = 1] {figs/data/RES_SF11_Pi-3.00_OS10_phaseOffset1_symmetries0_0.dat};
		\addlegendentry{SF${=}11$}
		% SF = 12
		\addplot[Set1-7-6, thick, solid, mark=diamond*, mark options={scale=1.3}] table[x index=0, y index = 1] {figs/data/RES_SF12_Pi-3.00_OS10_phaseOffset1_symmetries0_0.dat};
		\addlegendentry{SF${=}12$};

		%% No phase offset
		\addlegendimage{empty legend}
    \addlegendentry{\textbf{Non-PO:}} 
		\addplot[black, thick, dotted, mark=star, mark options={scale=0.6, solid}] table[x index=0, y index = 1] {figs/data/RES_SF7_Pi-3.00_OS10_phaseOffset0_symmetries0_0.dat};
		\addlegendentry{SF${=}7$}
		\addplot[black, thick, dotted, mark=pentagon*, mark options={scale=0.5, solid}] table[x index=0, y index = 1] {figs/data/RES_SF8_Pi-3.00_OS10_phaseOffset0_symmetries0_0.dat};
		\addlegendentry{SF${=}8$}
		\addplot[black, thick, dotted, mark=*, mark options={scale=0.6, solid}] table[x index=0, y index = 1] {figs/data/RES_SF9_Pi-3.00_OS10_phaseOffset0_symmetries0_0.dat};
		\addlegendentry{SF${=}9$}
		\addplot[black, thick, dotted, mark=square*, mark options={scale=0.6, solid}] table[x index=0, y index = 1] {figs/data/RES_SF10_Pi-3.00_OS10_phaseOffset0_symmetries0_0.dat};
		\addlegendentry{SF${=}10$}
		\addplot[black, thick, dotted, mark=triangle*, mark options={scale=0.6, solid}] table[x index=0, y index = 1] {figs/data/RES_SF11_Pi-3.00_OS10_phaseOffset0_symmetries0_0.dat};
		\addlegendentry{SF${=}11$}
		\addplot[black, thick, dotted, mark=diamond*, mark options={scale=0.6, solid}] table[x index=0, y index = 1] {figs/data/RES_SF12_Pi-3.00_OS10_phaseOffset0_symmetries0_0.dat};
		\addlegendentry{SF${=}12$}

		\addplot[black, ultra thick, solid, opacity=0.25] table[x index=0, y index = 1] {figs/data/RES_SF7_Pi-Inf_OS1_phaseOffset0_symmetries0_0.dat};
		\addplot[black, ultra thick, solid, opacity=0.25] table[x index=0, y index = 1] {figs/data/RES_SF8_Pi-Inf_OS1_phaseOffset0_symmetries0_0.dat};
		\addplot[black, ultra thick, solid, opacity=0.25] table[x index=0, y index = 1] {figs/data/RES_SF9_Pi-Inf_OS1_phaseOffset0_symmetries0_0.dat};
		\addplot[black, ultra thick, solid, opacity=0.25] table[x index=0, y index = 1] {figs/data/RES_SF10_Pi-Inf_OS1_phaseOffset0_symmetries0_0.dat};
		\addplot[black, ultra thick, solid, opacity=0.25] table[x index=0, y index = 1] {figs/data/RES_SF11_Pi-Inf_OS1_phaseOffset0_symmetries0_0.dat};
		\addplot[black, ultra thick, solid, opacity=0.25] table[x index=0, y index = 1] {figs/data/RES_SF12_Pi-Inf_OS1_phaseOffset0_symmetries0_0.dat};

	\end{semilogyaxis}

\end{tikzpicture}%
  \caption{Symbol error rate of the LoRa modulation under AWGN and same-SF interference for $\text{SF} \in \left\{7,\hdots,12\right\}$ and $P_I = {-}3$ dB.}
  \label{fig:serintphase}
\end{figure}
Using the Gaussian approximation, $P^{(I)}(\hat{s}{\neq} s)$ is
\begin{align}
  P^{(I)}(\hat{s}{\neq} s)  & = \frac{1}{N^3}\sum _{s_{I_1}=0}^{N{-}1}\sum _{s_{I_2}=0}^{N{-}1} \int _{0}^{N} Q\left(\frac{N-|h_I||R_{\max}|}{\sqrt{2\sigma^2}}\right)d \tau, \label{eq:approxint}
\end{align}
where $Q(\cdot)$ denotes the Q-function and the integral can be evaluated numerically by discretizing the interval $[0,N)$ with a step size $\epsilon$. In the AWGN-limited regime, the above approximation becomes inaccurate, since no single bin dominates the error rate. Let $P^{(N)}(\hat{s}\neq s)$ denote the SER under AWGN given in~\eqref{eq:formula_SER_only_noise_finalform}. Then, a final estimate of the SER that is more accurate also in the low SNR regime can be obtained as
\begin{align}
    P(\hat{s}{\neq} s)  & \approx P^{(N)}(\hat{s}{\neq} s) {+} \left(1{-}P^{(N)}(\hat{s}{\neq} s)\right)P^{(I)}(\hat{s}{\neq} s). \label{eq:approxfinal}
\end{align}

\section{Results} \label{sec:results}

In this section, we provide numerical results for the SER of a LoRa user with same-SF interference. In the remainder of this section, we use $\epsilon = \nicefrac{1}{10}$ to discretize the integral in~\eqref{eq:approxint}.

In Fig.~\ref{fig:serintphase}, we show the results of a Monte Carlo simulation for the SER of a LoRa user for all possible spreading factors $\text{SF} \in \left\{7,\hdots,12\right\}$, under the effect of same-SF interference with an SIR of $3$ dB (i.e., $P_I=-3$ dB) and AWGN. The SER when there is only AWGN is also included in the figure with thick transparent lines. We can clearly observe the strong impact of the interference on the SER when comparing to the case where there is only AWGN. The black dotted lines in the figure depict the SER when the relative phase offset between the interferer and the user $\omega$ is not taken into account in the Monte Carlo simulation. It is interesting to observe that $\omega$ does not seem to play an important role for the SER, which further justifies ignoring $\omega$ in the approximation of Section~\ref{sec:SER_approx}.

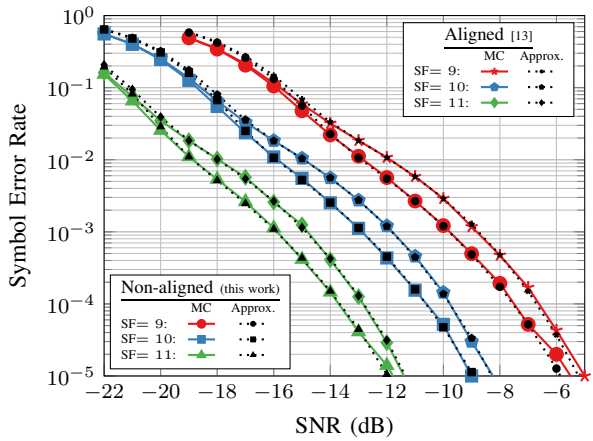
\begin{figure}[t]
  \centering
  \begin{tikzpicture}

	%\pgfplotsset{grid style={dashed}}
	\small

	\begin{semilogyaxis}[
		width = \figurewidth\columnwidth,
		height = \figureheight\columnwidth,
		xlabel = {SNR (dB)},
		ylabel = {Symbol Error Rate},
		label style={font=\small},
    tick label style={font=\footnotesize},
		ylabel near ticks,
		xlabel near ticks,
		xmin = -22, xmax = -5,
		ymin = 1e-5, ymax = 1,
		grid = both,
		legend image post style={scale=0.6},
		%legend style={at={(0.45,-0.25)},anchor=north,font=\tiny},
		%legend style={font=\tiny},
		%legend cell align={left},
		%legend columns={3},
		%transpose legend,
	]

		%%% Phase-aligned, chip-aligned %%%

		%% Monte Carlo simulation %%
		% SF = 9
		\addplot[Set1-7-1, thick, solid, mark=star, mark options={scale=1.2}] table[x index=0, y index = 1] {figs/data/RES_SF9_Pi-3.00_OS1_phaseOffset0_symmetries0_0.dat};
		\label{SF9MCPACA}
		%\addlegendentry{SF${=}9$}
		% SF = 10
		\addplot[Set1-7-2, thick, solid, mark=pentagon*, mark options={scale=1.2}] table[x index=0, y index = 1] {figs/data/RES_SF10_Pi-3.00_OS1_phaseOffset0_symmetries0_0.dat};
		\label{SF10MCPACA}
		%\addlegendentry{SF${=}10$}
		% SF = 11
		\addplot[Set1-7-3, thick, solid, mark=diamond*, mark options={scale=1.3}] table[x index=0, y index = 1] {figs/data/RES_SF11_Pi-3.00_OS1_phaseOffset0_symmetries0_0.dat};
		\label{SF11MCPACA}
		%\addlegendentry{SF${=}11$};

		%% Approximation %%
		\addplot[black, thick, dotted, mark=star, mark options={scale=0.6, solid}] table[x index=0, y index = 1] {figs/data/APP_SF9_Pi-3.00_OS1.dat};
		\label{SF9APPACA}
		%\addlegendentry{SF${=}9$}
		% SF = 10
		\addplot[black, thick, dotted, mark=pentagon*, mark options={scale=0.6, solid}] table[x index=0, y index = 1] {figs/data/APP_SF10_Pi-3.00_OS1.dat};
		\label{SF10APPACA}
		%\addlegendentry{SF${=}10$}
		% SF = 11
		\addplot[black, thick, dotted, mark=diamond*, mark options={scale=0.6, solid}] table[x index=0, y index = 1] {figs/data/APP_SF11_Pi-3.00_OS1.dat};
		\label{SF11APPACA}
		%\addlegendentry{SF${=}11$};

		%%% Non-phase-aligned, non-chip-aligned %%%

		%% Monte Carlo simulation %%
		% SF = 9
		\addplot[Set1-7-1, thick, solid, mark=*, mark options={scale=1.2}] table[x index=0, y index = 1] {figs/data/RES_SF9_Pi-3.00_OS10_phaseOffset1_symmetries0_0.dat};
		\label{SF9MCPNCN}
		%\addlegendentry{SF${=}9$}
		% SF = 10
		\addplot[Set1-7-2, thick, solid, mark=square*, mark options={scale=1.05}] table[x index=0, y index = 1] {figs/data/RES_SF10_Pi-3.00_OS10_phaseOffset1_symmetries0_0.dat};
		\label{SF10MCPNCN}
		%\addlegendentry{SF${=}10$}
		% SF = 11
		\addplot[Set1-7-3, thick, solid, mark=triangle*, mark options={scale=1.3}] table[x index=0, y index = 1] {figs/data/RES_SF11_Pi-3.00_OS10_phaseOffset1_symmetries0_0.dat};
		\label{SF11MCPNCN}
		%\addlegendentry{SF${=}11$};

		%% Approximation %%
		% SF = 9
		\addplot[black, thick, dotted, mark=*, mark options={scale=0.6, solid}] table[x index=0, y index = 1] {figs/data/APP_SF9_Pi-3.00_OS3.dat};
		\label{SF9APPNCN}
		%\addlegendentry{SF${=}9$}
		% SF = 10
		\addplot[black, thick, dotted, mark=square*, mark options={scale=0.6, solid}] table[x index=0, y index = 1] {figs/data/APP_SF10_Pi-3.00_OS3.dat};
		\label{SF10APPNCN}
		%\addlegendentry{SF${=}10$}
		% SF = 11
		\addplot[black, thick, dotted, mark=triangle*, mark options={scale=0.6, solid}] table[x index=0, y index = 1] {figs/data/APP_SF11_Pi-3.00_OS3.dat};
		\label{SF11APPNCN}
		%\addlegendentry{SF${=}11$};

		% Draw first "Legend" node using a left justified shortstack, position using relative axis coordinates
		\node [draw,fill=white,inner sep=2pt] at (rel axis cs: 0.2,0.15) {
		\tiny
		\begin{tabular}{lcc}
			\multicolumn{3}{c}{\scriptsize\underline{Non-aligned {\tiny(this work)}}} \\
								& MC 								& Approx. \\
			SF$=9$:	 	& \ref{SF9MCPNCN} 	&	\ref{SF9APPNCN} \\
			SF$=10$: 	& \ref{SF10MCPNCN} & \ref{SF10APPNCN} \\
			SF$=11$: 	& \ref{SF11MCPNCN} & \ref{SF11APPNCN}
		\end{tabular}};

		\node [draw,fill=white,inner sep=2pt] at (rel axis cs: 0.8,0.85) {
		\tiny
		\begin{tabular}{lcc}
			\multicolumn{3}{c}{\scriptsize\underline{Aligned~{\tiny\cite{Elshabrawy2018b}}} } \\
								& MC 								& Approx. \\
			SF$=9$:	 	& \ref{SF9MCPACA} 	&	\ref{SF9APPACA} \\
			SF$=10$: 	& \ref{SF10MCPACA} & \ref{SF10APPACA} \\
			SF$=11$: 	& \ref{SF11MCPACA} & \ref{SF11APPACA}
		\end{tabular}};

	\end{semilogyaxis}

\end{tikzpicture}%
  \caption{Symbol error rate of the LoRa modulation under AWGN and same-SF interference for $\text{SF} \in \left\{9,10,11\right\}$ and $P_I = {-}3$ dB.}
  \label{fig:serint}
\end{figure}

In Fig.~\ref{fig:serint}, we show the results of a Monte Carlo simulation for the SER of a LoRa user for $\text{SF} \in \left\{9,10,11\right\}$ using the chip-aligned model of~\cite{Elshabrawy2018b} and the  model we described in this work, as well as the corresponding approximations in~\cite{Elshabrawy2018b} and~\eqref{eq:approxfinal}, respectively. We observe that there is a significant difference of approximately $1$~dB between the two models. The chip-aligned model of~\cite{Elshabrawy2018b} is pessimistic in the computation of the SER. Finally, we clearly observe that the low-complexity computation of~\eqref{eq:ser_full} using the approximation derived in Section~\ref{sec:SER_approx} is very accurate.

\section{Conclusion} \label{sec:conclusion}
In this work, we introduced a LoRa interference model where the interference is neither chip- nor phase-aligned with the LoRa signal-of-interest and we derived a corresponding expression for the SER. Moreover, we derived a low-complexity approximation for the SER and we showed that ignoring the non-integer time offsets overestimates the error rate by $1$~dB.

\bibliographystyle{IEEEtran}
\bibliography{IEEEabrv,refs}

\begin{thebibliography}{10}
\providecommand{\url}[1]{#1}
\csname url@samestyle\endcsname
\providecommand{\newblock}{\relax}
\providecommand{\bibinfo}[2]{#2}
\providecommand{\BIBentrySTDinterwordspacing}{\spaceskip=0pt\relax}
\providecommand{\BIBentryALTinterwordstretchfactor}{4}
\providecommand{\BIBentryALTinterwordspacing}{\spaceskip=\fontdimen2\font plus
\BIBentryALTinterwordstretchfactor\fontdimen3\font minus
  \fontdimen4\font\relax}
\providecommand{\BIBforeignlanguage}[2]{{%
\expandafter\ifx\csname l@#1\endcsname\relax
\typeout{** WARNING: IEEEtran.bst: No hyphenation pattern has been}%
\typeout{** loaded for the language `#1'. Using the pattern for}%
\typeout{** the default language instead.}%
\else
\language=\csname l@#1\endcsname
\fi
#2}}
\providecommand{\BIBdecl}{\relax}
\BIBdecl

\bibitem{SX127x}
\BIBentryALTinterwordspacing
{Semtech Corporation}, ``{SX1272/73}--860 {MHz} to 1020 {MHz} low power long
  range transceiver.'' [Online]. Available:
  \url{https://www.semtech.com/uploads/documents/sx1272.pdf}
\BIBentrySTDinterwordspacing

\bibitem{Knight2016}
M.~Knight and B.~Seeber, ``Decoding {LoRa}: Realizing a modern {LPWAN} with
  {SDR},'' in \emph{GNU Radio Conf.}, Sep. 2016.

\bibitem{Robyns2018}
P.~Robyns, P.~Quax, W.~Lamotte, and W.~Thenaers, ``A multi-channel software
  decoder for the {LoRa} modulation scheme,'' in \emph{Int. Conf. on Internet
  of Things, Big Data and Security {(IoTBDS)}}, Mar. 2018.

\bibitem{Vangelista2017}
L.~{Vangelista}, ``Frequency shift chirp modulation: The {LoRa} modulation,''
  \emph{IEEE Signal Process. Lett.}, vol.~24, no.~12, pp. 1818--1821, Dec.
  2017.

\bibitem{Ghanaatian2019}
R.~Ghanaatian, O.~Afisiadis, M.~Cotting, and A.~Burg, ``{LoRa} digital receiver
  analysis and implementation,'' \emph{ArXiv e-prints}, Feb. 2019,
  \url{https://arxiv.org/abs/1811.04146}.

\bibitem{Bor2016}
M.~C. Bor, U.~Roedig, T.~Voigt, and J.~M. Alonso, ``Do {LoRa} low-power
  wide-area networks scale?'' in \emph{ACM Int. Conf. on Modeling, Analysis and
  Simulation of Wireless and Mobile Systems}, ser. MSWiM '16.\hskip 1em plus
  0.5em minus 0.4em\relax New York, NY, USA: ACM, 2016, pp. 59--67.

\bibitem{Haxhibeqiri2018}
J.~Haxhibeqiri, E.~De~Poorter, I.~Moerman, and J.~Hoebeke, ``A survey of
  {LoRaWAN} for {IoT}: From technology to application,'' \emph{Sensors},
  vol.~18, no.~11, 2018.

\bibitem{Orfanidis2017}
C.~{Orfanidis}, L.~M. {Feeney}, M.~{Jacobsson}, and P.~{Gunningberg},
  ``Investigating interference between {LoRa} and {IEEE} 802.15.4g networks,''
  in \emph{IEEE Int. Conf. on Wireless and Mobile Computing, Networking and
  Communications (WiMob)}, Oct. 2017, pp. 1--8.

\bibitem{Reynders2016}
B.~{Reynders}, W.~{Meert}, and S.~{Pollin}, ``Range and coexistence analysis of
  long range unlicensed communication,'' in \emph{Int. Conf. on
  Telecommunications (ICT)}, May 2016, pp. 1--6.

\bibitem{Croce2018}
D.~{Croce}, M.~{Gucciardo}, S.~{Mangione}, G.~{Santaromita}, and
  I.~{Tinnirello}, ``Impact of {LoRa} imperfect orthogonality: Analysis of
  link-level performance,'' \emph{IEEE Commun. Lett.}, vol.~22, no.~4, pp.
  796--799, Apr. 2018.

\bibitem{Feltrin2018}
L.~{Feltrin}, C.~{Buratti}, E.~{Vinciarelli}, R.~{De Bonis}, and R.~{Verdone},
  ``{LoRaWAN}: Evaluation of link- and system-level performance,'' \emph{IEEE
  Internet of Things Journal}, vol.~5, no.~3, pp. 2249--2258, Jun. 2018.

\bibitem{Elshabrawy2018}
T.~Elshabrawy and J.~Robert, ``Closed-form approximation of {LoRa} modulation
  {BER} performance,'' \emph{IEEE Commun. Lett.}, vol.~22, no.~9, pp.
  1778--1781, Sep. 2018.

\bibitem{Elshabrawy2018b}
T.~{Elshabrawy} and J.~{Robert}, ``Analysis of {BER} and coverage performance
  of {LoRa} modulation under same spreading factor interference,'' in
  \emph{IEEE Int. Symp. on Personal, Indoor and Mobile Radio Communications
  (PIMRC)}, Sep. 2018.

\bibitem{Elshabrawy2019}
T.~{Elshabrawy} and J.~{Robert}, ``Capacity planning of {LoRa} networks with
  joint noise-limited and interference-limited coverage considerations,''
  \emph{IEEE Sensors Journal}, pp. 1--1, Feb. 2019.

\bibitem{David2003}
H.~A. David and H.~N. Nagaraja, \emph{Order Statistics}.\hskip 1em plus 0.5em
  minus 0.4em\relax Wiley, 2003.

\bibitem{Adelantado2017}
F.~{Adelantado}, X.~{Vilajosana}, P.~{Tuset-Peiro}, B.~{Martinez},
  J.~{Melia-Segui}, and T.~{Watteyne}, ``Understanding the limits of
  {LoRaWAN},'' \emph{IEEE Commun. Mag.}, vol.~55, no.~9, pp. 34--40, Sep. 2017.

\bibitem{Goursaud2015}
C.~Goursaud and J.-M. Gorce, ``{Dedicated networks for {IoT} : {PHY} / {MAC}
  state of the art and challenges},'' \emph{{EAI endorsed trans. on Internet of
  Things}}, Oct. 2015.

\end{thebibliography}

\end{document}